\documentclass[twocolumn,showpacs,superscriptaddress,aps,pra]{revtex4-1}
\usepackage{epsfig}
\usepackage{subfigure}
\usepackage{amssymb}
\usepackage{graphicx}
\usepackage{color}
\usepackage{amsfonts}
\usepackage{amscd}
\usepackage{hyperref}
\usepackage{amsmath}
\usepackage{epstopdf}

\usepackage{amsmath}
\usepackage{graphicx}
\usepackage{txfonts}
\usepackage{textcomp}

\begin{document}

\title{Robust stimulated Raman shortcut-to-adiabatic passage by invariant-based optimal control}

\author{Xue-Ke Song}
\affiliation{School of Physics and Material Science, Anhui University, Hefei 230601, China}
\affiliation{Shenzhen Institute for Quantum Science and Engineering and Department of Physics, Southern University of Science and Technology, Shenzhen 518055, China}

\author{Fei Meng}
\affiliation{Shenzhen Institute for Quantum Science and Engineering and Department of Physics, Southern University of Science and Technology, Shenzhen 518055, China}
\affiliation{QICI Quantum Information and Computation Initiative, Department of Computer Science, The University of Hong Kong, Pokfulam Road, Hong Kong SAR, China}

\author{Bao-Jie Liu}
\affiliation{Shenzhen Institute for Quantum Science and Engineering and Department of Physics, Southern University of Science and Technology, Shenzhen 518055, China}

\author{Dong Wang}
\email{dwang@ahu.edu.cn}
\affiliation{School of Physics and Material Science, Anhui University, Hefei 230601, China}

\author{Liu Ye}
\email{yeliu@ahu.edu.cn}
\affiliation{School of Physics and Material Science, Anhui University, Hefei 230601, China}

\author{Man-Hong Yung}
\email{yung@sustech.edu.cn}
\affiliation{Shenzhen Institute for Quantum Science and Engineering and Department of Physics, Southern University of Science and Technology, Shenzhen 518055, China}
\affiliation{Shenzhen Key Laboratory of Quantum Science and Engineering, Southern University of Science and Technology, Shenzhen, 518055, China}

\date{\today }

\begin{abstract}
The stimulated Raman adiabatic passage (STIRAP) shows an efficient technique that accurately transfers population between two discrete quantum states
with the same parity, in three-level quantum systems based on adiabatic evolution. This technique has widely
theoretical and experimental applications in many fields of physics, chemistry, and beyond. Here, we present a generally robust approach to speed up STIRAP with invariant-based shortcut to adiabaticity. By controlling the dynamical process, we inversely design a family of Hamiltonians that can realize fast and accurate population transfer from the first to the third level, while the systematic error is largely suppressed in general.
Furthermore, a detailed trade-off relation between the population of the intermediate state and the amplitudes of
Rabi frequencies in the transfer process is illustrated.
These results provide an optimal route toward manipulating the evolution of three-level quantum systems in future quantum information processing.
\end{abstract}
\maketitle

\section{Introduction}

The accurate manipulation of quantum systems with high-fidelity is of great importance in atomic, molecular, optical
physics and chemistry~\cite{molecular1,molecular2,molecular3,molecular4}.
In particular, stimulated Raman adiabatic passage (STIRAP)~\cite{molecular4, stirap} is a technique that
can control preparation of the internal state and the dynamics of three-level quantum systems, so that an
efficient and selective population transfer between quantum states is implemented. Population transfer by STIRAP has two
distinct advantages: (i) it is robust against loss due to spontaneous emission from the intermediate state;  (ii)
it is insensitive to small variations of laser intensity, the duration, and the delay of
pulses when the adiabatic condition is fulfilled. However, the adiabatic evolution in STIRAP requires a long run time
for the desired parametric control, which makes the systems vulnerable to open system effects~\cite{vitanov2001}.
On the other hand, resonant pulses can achieve fast quantum state engineering, but they are sensitive to the
parameter fluctuations that may lead to loss of fidelity. Therefore, implementing an effective method that
combines both the the advantages of adiabatic control and resonant interactions, which realizes fast
and robust quantum state control, is a hot topic in recent years.

Shortcuts to adiabaticity (STA), introduced by Chen \emph{et al.} in 2010~\cite{chen2010}, are optimal control protocols that can speed
up a quantum adiabatic process through a non-adiabatic route. They offer alternative
fast and accurate processes which can reproduce the same final populations, or even the
same final state in a shorter time, compared to the adiabatic evolution~\cite{STAreview2013,STAreview2019}.
Recently, STA have been widely used to implement rapid and robust quantum information processing in theory, such
as quantum state transfer~\cite{chen2010three,chen2012,giannelli2014,chenxia2014,song2016,li2016,wu2020},
quantum gates~\cite{chen2015,song2016a,palmero2017,liu2019,du2019,wu2019,lisai2020,lv2020}, and generation of quantum cat
state~\cite{hatomura2019,palmero2019,chen2020}, etc. For instance, in 2012, Chen \emph{et al.}~\cite{chen2012} used the invariant-based
shortcut to accelerate the STIRAP to realize population transfers in a short time. In 2016, Li \emph{et al.}~\cite{li2016} achieved
fast and accurate population transfer in three-level quantum systems with a stimulated Raman
shortcut-to-adiabatic passage (STISRAP), based on counterdiabatic driving. In 2019, Liu \emph{et al.}~\cite{liu2019}
introduced a NHQC+ protocol, which can incorporate STA to construct a variety of extensible nonadiabatic geometric gates that are
robust against several types of noises.
Experimentally, efforts have also been made for investigating the applications of STA in many quantum
systems~\cite{bason2012,du2016,zhou2017,yan2019,Niu2019,wang2019,vepsalainen2019,kolbl2019}. In 2016, Du \emph{et al.}~\cite{du2016} demonstrated a fast and high-fidelity STISRAP to speed-up the conventional "slow" STIRAP. In 2019, K\"{o}lbl
\emph{et al.}~\cite{kolbl2019} employed the STA state transfer protocol to realize high-fidelity, reversible initialization of
individual dressed states in a closed-contour, coherent driving of a single spin system.

Noises, errors, and fluctuations will decrease the accuracy
and efficiency of the control of quantum states. In particular, some systematic errors
cannot be avoided. For example, atoms take different Rabi frequencies induced by different positions
and fields~\cite{ruschhaupt2012}, and spin systems get different shifts of the amplitude of magnetic field caused by imperfections~\cite{yu2018}. To suppress these errors, the invariant-based inverse engineering~\cite{ruschhaupt2012,yu2018,lu2013,song2017,santos2018,laforgue2019} is an effective method. In 2012, Ruschhaupt \emph{et al.}~\cite{ruschhaupt2012}
explored the stability of different shortcut protocols with respect to systematic errors in a generic two-level system. In 2018,
Yu \emph{et al.}~\cite{yu2018} proposed a STA approach against systematic errors, to achieve robust spin flip and to generate the entangled Bell state. In 2019, Laforgue \emph{et al.}~\cite{laforgue2019}
showed an exact and robust ultrahigh-fidelity population transfer on a three-level
quantum system, using the Lewis-Riesenfeld (LR) method.

In this paper, we employ the invariant-based inverse engineering to realize a fast and robust three-level
population transfer in a general setting. Specifically, we analytically find one solution of dynamic parameters for the optimal shortcut
with zero systematic error sensitivity (SES). More importantly, we show that the SES is
always very small for a family of solutions of a given functional form.
Compared to the previous protocols, the optimal shortcuts can realize a perfect population transfer in a more rapid and accurate way, while the time evolution of Rabi frequencies do not diverge, which implies that the frequencies can be easily implemented in the laboratory.
Moreover, in the case of negligible SES, a close relation between the amplitude of Rabi frequencies and population of intermediate state
are formed; the smaller the population of the intermediate state is, the larger the
maximal amplitudes of Rabi frequencies are. This offers alternative routes for achieving high-fidelity state transfer of
three-level quantum systems.

This paper is organized as follows: In Sec. \ref{sec2}, we present the
basic principle of STISRAP by using invariant-based inverse engineering.
In Sec. \ref{sec3}, we combine the concept of STA with perturbation theory based on Dyson series
to design optimal STISRAP against systematic errors. In Sec. \ref{sec4}, we study the trade-off
between the maximal amplitude of Rabi frequencies and maximal population of intermediate state for the optimal shortcut.
In Sec. \ref{sec5}, we compare the robustness of optimal invariant shortcut with the original one.
A summary is provided in Sec. \ref{sec6}.

\section{Stimulated Raman shortcut-to-adiabatic passage by invariant-based optimal control}
\label{sec2}

We denote the three orthogonal quantum states of a three-level system by $\{\left|1\right\rangle,\left|2\right\rangle,\left|3\right\rangle\}$,
with $\left|2\right\rangle$ being the intermediate state, and $\left|1\right\rangle$ and $\left|3\right\rangle$
being two lower-energy states which represent the initial and target state respectively.
The STIRAP is an efficient method that can realize the selective population
transfer between the state $\left|1\right\rangle$ and the state $\left|3\right\rangle$ by controlling the system to evolve
along the adiabatic dark state. It involves a three-state,
two-photon Raman process with counterintuitive pulse orders. More specifically, the three-level quantum system
first interacts with the Stokes laser, which links the state $\left|3\right\rangle$ with the state $\left|2\right\rangle$,
and then it interacts with the pump laser, which leads to the
transition between the state $\left|1\right\rangle$ and the state $\left|2\right\rangle$.
Meanwhile, to make sure that the state
vector of the system adiabatically follows the evolution of the
dark state, there must be a sufficient coupling (an appropriate overlap) between the two pulses.
The Hamiltonian, under the
rotating wave approximation with $\hbar=1$, takes the form of
\begin{eqnarray}    
H(t)=\frac{1}{2}\left(\begin{array}{ccc}
0 & \Omega_{p}(t) & 0\\
\Omega_{p}(t) & 0 & \Omega_{s}(t)\\
0 & \Omega_{s}(t) & 0
\end{array}\right),\label{hamil}
\end{eqnarray}
where Rabi frequencies $\Omega_{p}(t)$ and $\Omega_{s}(t)$ describe the
interactions with the pump and Stokes fields. The Hamiltonian can
be found, for example, in quantum-dots, or superconducting quantum
systems. It can be mapped to the Hamiltonian of a spin-1 system, which is
\begin{eqnarray}    
H\left(t\right)=\frac{1}{2}\left(\Omega_{p}(t)J_{1}+\Omega_{s}(t)J_{2}\right),
\end{eqnarray}
where $J_{\nu}$$(\upsilon=1,2,3)$ are spin-1 generator matrices,
fulfilling the SU(2) algebra $[J_{\mu},J_{\nu}]=iJ_{\gamma}\varepsilon_{\mu\nu\gamma}$.
For the Hamiltonian $H\left(t\right)$, there exists another explicitly time-dependent nontrivial
Hermitian operator $I(t)$, a dynamical invariant, satisfying the equation
\begin{eqnarray}    
\frac{dI}{dt}=\frac{1}{i\hbar}\left[I,H\right]+\frac{\partial I}{\partial t}=0.\label{dynamical}
\end{eqnarray}
The invariant can be constructed by the superposition of the three group generators of spin-1 matrices with parameters $\theta$ and $\beta$,
\begin{eqnarray}    
I\left(t\right)=\frac{1}{2}B_{0}\left(\begin{array}{ccc}
0 & \cos\theta\sin\beta & -i\sin\theta\\
\cos\theta\sin\beta & 0 & \cos\theta\cos\beta\\
i\sin\theta & \cos\theta\cos\beta & 0
\end{array}\right),
\end{eqnarray}
where $B_{0}$ is a constant magnitude of magnetic field, guaranteeing
the same energy dimension as $H(t)$. The eigenvectors of the invariant are
\begin{eqnarray}    
\left|\phi_{0}\left(t\right)\right\rangle =\left(\begin{array}{c}
\cos\theta\cos\beta\\
-i\sin\theta\\
-\cos\theta\sin\beta
\end{array}\right), \label{eigen0}
\end{eqnarray}
\begin{eqnarray}    
\left|\phi_{1}\left(t\right)\right\rangle =\frac{1}{\sqrt{2}}\left(\begin{array}{c}
\sin\theta\cos\beta+i\sin\beta\\
i\cos\theta\\
-\sin\theta\sin\beta+i\cos\beta
\end{array}\right),\label{eigen1}
\end{eqnarray}
\begin{eqnarray}    
\left|\phi_{2}\left(t\right)\right\rangle =\frac{1}{\sqrt{2}}\left(\begin{array}{c}
\sin\theta\cos\beta-i\sin\beta\\
i\cos\theta\\
-\sin\theta\sin\beta-i\cos\beta
\end{array}\right),\label{eigen2}
\end{eqnarray}
with corresponding eigenvalues $\lambda_{0}=0$ and $\lambda_{1,2}=\pm B_{0}/2$. By solving the
dynamical equation of Eq.~(\ref{dynamical}), one has
\begin{eqnarray}    
\dot{\theta}=\frac{1}{2}\left(\Omega_{p}\cos\beta-\Omega_{s}\sin\beta\right),
\end{eqnarray}
\begin{eqnarray}    
\dot{\beta}=\frac{1}{2}\tan\theta\left(\Omega_{s}\cos\beta+\Omega_{p}\sin\beta\right).
\end{eqnarray}
The Rabi frequencies $\Omega_{p}$ and $\Omega_{s}$ can be solved as
\begin{eqnarray}    
\Omega_{p}=2\left(\dot{\beta}\cot\theta\sin\beta+\dot{\theta}\cos\beta\right), \label{omegap}
\end{eqnarray}
\begin{eqnarray}    
\Omega_{s}=2\left(\dot{\beta}\cot\theta\cos\beta-\dot{\theta}\sin\beta\right).\label{omegas}
\end{eqnarray}
According to the theory of Lewis and Riesenfeld~\cite{lewis1969}, the general solution of the Schrodinger
equation can be written as the superposition of the three  orthogonal ``dynamical modes'' of the invariant $I(t)$, that is,
\begin{eqnarray}    
\left|\Psi\left(t\right)\right\rangle =\sum_{n}c_{n}e^{i\gamma_{(n)}}\left|\phi_{n}\left(t\right)\right\rangle ,
\end{eqnarray}
where $c_{n}$ is a \emph{time-independent} constant. The wave function $|{\Psi(t)}\rangle$ is a supposition of time-dependent eigenstates $|{\phi_n(t)}\rangle$ with \emph{constant} amplitudes (up to the LR phase factor $e^{i \gamma_{(n)}})$, i.e., the dynamics follows the eigenstates of the invariant $I(t)$
without any approximation. In this sense, the invariant-based shortcut method breaks the speed limitation imposed by the adiabatic condition and therefore allows us
to design the dynamics as fast as we wish, by choosing an appropriate time-dependent invariant. For this three-level system with the Hamiltonian $H(t)$ and the  time-dependent invariant $I(t)$, we can solve the LR phases $\gamma_{(n)}$ as
\begin{eqnarray}    
\gamma_{(0)}=0,
\end{eqnarray}
\begin{eqnarray}    
\gamma_{(1,2)}=\!\mp\!\!\!\int_{0}^{t}\!\!dt'\left[\dot{\beta}\sin\theta+\frac{1}{2}\left(\Omega_{p}\sin\beta\!+\!\Omega_{s}\cos\beta\right)\cos\theta\right]. \label{gamma}
\end{eqnarray}
Then, three orthogonal dynamical solutions can be constructed as
\begin{eqnarray}    
|\psi_{n}(t)\rangle=e^{i\gamma_{(n)}(t)}\left|\phi_{n}\left(t\right)\right\rangle,
\end{eqnarray}
and $\left\langle\psi_{n}(t)|\psi_{m}(t)\right\rangle=0$ is valid for all times when $m\neq n$,
where $m,n=0,1,2$.
Substituting Eqs.~(\ref{omegap}) and ~(\ref{omegas}) into Eqs.~(\ref{gamma}), we get a constraint for the parameters $\theta$ and $\beta$ as
\begin{eqnarray}    
\sin\theta=\dot{\beta}/\dot{\gamma},
\end{eqnarray}
where $\gamma=\gamma_{(2)}=-\gamma_{(1)}$.

Now we are ready to apply the inverse engineering to achieve the fast quantum driving along the eigenstate of invariant
$\left|\phi_{0}\left(t\right)\right\rangle$ in a given
time $T$. If
we want to inverse the population from $\left|1\right\rangle $ to
$\left|3\right\rangle$ along this eigenstate, the boundary conditions
are given by
\begin{eqnarray}    
\Omega_{p}\left(0\right)=\Omega_{s}\left(T\right)&=&0,
\nonumber \\
\;\;\;\;\beta\left(0\right)=0,\;\;\;\beta\left(T\right)&=&\pi/2,
\nonumber \\
\;\;\;\;\dot{\beta}\left(0\right)=\dot{\beta}\left(T\right)&=&0,
\nonumber \\
\!\!\!\!\!\!\theta\left(0\right)=\theta\left(T\right)&=&k\pi,
\nonumber \\
\!\!\!\!\!\!\dot{\theta}\left(0\right)=\dot{\theta}\left(T\right)&=&0. \label{boundary}
\end{eqnarray}
Once the parameters $\beta$ and $\theta$ are determined according to the boundary conditions, the Rabi frequencies for a rapid and high-fidelity STISRAP
are found from Eqs.~(\ref{omegap}) and ~(\ref{omegas}).

\section{Optimal STISRAP based on the Dyson perturbative series}
\label{sec3}

Here we consider systematic errors as the ``perturbing'' term
representing a weak disturbance to the system. Combining invariant-based shortcuts and  perturbation theory,
we can get approximate solutions of the Schr\"{o}dinger equation and design a robust quantum state
engineering against such sources of errors. The perturbing Hamiltonian is assumed to take the form
\begin{eqnarray}    
H'\left(t\right)=\frac{1}{2}\left(\begin{array}{ccc}
0 & \lambda\Omega_{p} & 0\\
\lambda\Omega_{p} & 0 & \lambda\Omega_{s}\\
0 & \lambda\Omega_{s} & 0
\end{array}\right), \label{pHamil}
\end{eqnarray}
that is, $H'=\lambda H$, where $\lambda$ is small time-independent and dimensionless parameter, thereby implying that
the Rabi frequencies have simultaneously a small shift, i.e.,  $\lambda\Omega_{p}$ and $\lambda\Omega_{s}$.

We use the technique of the Dyson series
to expand the wave function up to the second-order $O\left(\lambda^{2}\right)$,
\begin{eqnarray} 
&&\hspace{-1cm}|\psi(T)\rangle
= |\psi_{0}(T)\rangle -\frac{i}{\hbar}\int_{0}^{T}dt\hat{U}_{0}(t_{f},t)H'(t)\left|\psi_{0}(t)\right\rangle
\nonumber\\
&&\hspace{-1cm}-\frac{1}{\hbar^{2}}\!\int_{0}^{T}\!\!\!dt\!\!\!\int_{0}^{t}\!\!\!dt'\hat{U}_{0}(T,t)H'(t)\hat{U}_{0}(t,t')H'(t')\left|\psi_{0}(t')\right\rangle
+\cdot\cdot\cdot,
\end{eqnarray}
where $\left|\psi_{0}\left(t\right)\right\rangle $ is the unperturbed
solution and $\hat{U}_{0}\left(s,t\right)=\sum_{n}\left|\psi_{n}\left(s\right)\right\rangle \left\langle \psi_{n}\left(t\right)\right|$ is
the unperturbed time evolution operator.
We assume that the error-free $\left(\lambda=0\right)$ scheme works
perfectly, i.e., the system evolves exactly along the state $\left|\psi_{0}\left(t\right)\right\rangle$
from the initial state $\left|\psi_{0}\left(0\right)\right\rangle =\left|1\right\rangle$ to the final one
$\left|\psi_{0}\left(T\right)\right\rangle =\left|3\right\rangle$, up to a global phase factor.
Then the population of the state $\left|3\right\rangle$ at time $T$ is
\begin{eqnarray}    
\!\!P_{3}\!=\!\left|\left\langle \psi_{0}\left(T\right)\right.\!\!\left|\psi\left(T\right)\right\rangle \right|^{2}\!=\!\!1\!\!-\!\!\underset{n\neq0}{\sum}\!\left|\int_{0}^{T}\!\!\!dt\left\langle \psi_{0}\!\left(t\right)\right|H'\!\!\left(t\right)\!\left|\psi_{n}\!\left(t\right)\right\rangle\right|^{2}\!\!\!. \label{exicted}
\end{eqnarray}
Here the SES is defined as
\begin{eqnarray}    
q_{s}\coloneqq-\frac{1}{2}  \frac{\partial^{2}P_{3}(\lambda)}{\partial \lambda^{2}}\Bigg|_{\lambda=0}=
-\frac{\partial P_{3}(\lambda)}{\partial(\lambda^{2})}\Bigg|_{\lambda=0} ,
\end{eqnarray}
which is a dimensionless quantity that indicates the change of the population of state $|3\rangle$ compared to $\lambda^2$. For example, if $q_s = \epsilon$, then approximately, the population of $|3\rangle$ at $T$ is $1-\epsilon \lambda^2$. Therefore, the smaller the values of $q_{s}$ is, the more robust  that
the protocol is against these noises. From Eqs.~(\ref{pHamil}) and ~(\ref{exicted}), using
the eigenvectors of the invariant in Eqs.~(\ref{eigen0}),~(\ref{eigen1}), and~(\ref{eigen2}),
we find
\begin{eqnarray}    
q_{s}=\left|\int_{0}^{T}dte^{-i\gamma}\left(-i\overset{\cdot}{\theta}-\overset{\cdot}{\beta}\cos\theta\right)\right|^{2}.
\end{eqnarray}
We aim to design an "optimal" shortcut scheme with zero SES by setting
some particular conditions on the parameters $\theta$, $\beta$, and $\gamma$.
To this end, we do not need to find the complete set of general
solutions, but only need to find one solution for the parameters that satisfies
the boundary conditions. And for practical consideration, the functional forms of $\theta$ and $\beta$ should
be as simple as possible.

To begin with, let us try to simplify the boundary conditions by choosing an
appropriate functional form of $\theta$. Setting
\begin{eqnarray}    
\theta\left(t\right)=\frac{B}{2}\left[1-\cos\left(\frac{2\pi t}{T}\right)\right],\label{theta}
\end{eqnarray}
 the boundary conditions for $\theta$ are automatically satisfied,
\begin{eqnarray}    
\theta\left(0\right)&=&0,\;\;\;\theta\left(T\right)=0,
\nonumber \\
\theta\left(T/2\right)&=&B,\;\;\;\dot{\theta}\left(0\right)=\dot{\theta}\left(T\right)=0.\label{boun1}
\end{eqnarray}
Here $B$ is an arbitrary dimensionless constant which is called the population parameter, because $\sin^2 B$ represents the maximal value of population
of the intermediate state during the dynamic evolution. In practice, a higher value of $B$ will result in larger errors by the spontaneous emission of the intermediate state $|{2}\rangle$. Therefore, in designing the inverse-engineering, we should keep $B$ small, which guarantees that the population of the intermediate state can take a relatively small and controllable value in the transfer process. In this sense, we require that $B<\pi/2$.

Having assumed the functional form of $\theta$, the original boundary conditions imply that we should find a solution of $\beta\left(t\right)$ and $\gamma\left(t\right)$ such that
\begin{eqnarray}    
\beta\left(0\right)=0,\;\;\beta\left(T\right)=\frac{\pi}{2},\;\;\sin\theta=\frac{\dot{\beta}}{\dot{\gamma}}.\label{boun2}
\end{eqnarray}

Since the parameter $\theta$ is strictly increasing with respect
to $t$ for $t\in [0, T/2]$ and strictly decreasing for $t \in [T/2, T]$. By the inverse function theorem, we have
\begin{eqnarray}    
t_1\left(\theta\right)=\frac{T}{2\pi}\arccos\left(1-\frac{2\theta}{B}\right), \;\; \theta \in [0, B],
\end{eqnarray}
as the inverse function of $\theta(t)$ for $t\in [0, T/2]$ and $t_2(\theta) = T - t_1 (\theta)$ as the inverse function for $t\in [T/2,T]$.
This means that we can further simplify the boundary condition for the problem by setting $t=t_{1,2}\left(\theta\right)$
for the two time interval $t\in [0,T/2]$ and $t\in [T/2,T]$ respectively. In this way, the variable $t$ is eliminated and the problem can be further simplified as: $\tilde{b}_i\left(\theta\right)\equiv\beta\left(t_i\left(\theta\right)\right)$ and
$\tilde{\gamma}_i\left(\theta\right)\equiv\gamma\left(t_i\left(\theta\right)\right)$ for $i=1,2$, such
that
\begin{eqnarray}    
\tilde{b}_1(0)&=&0,\;\;\;\;\;\;\;\tilde{b}_1(B)=\tilde{b}_2(B)=A,
\nonumber \\
\tilde{b}_2(0)&=&\frac{\pi}{2},\;\;\;\;\;\;\;\sin\theta=\frac{\tilde{b}'_i}{\tilde{\gamma'}_i}.\label{gama}
\end{eqnarray}
for $i=1,2$, where $\tilde{b}'_i=\frac{d}{d\theta}\tilde{b}_i$, $\tilde{\gamma}'_i=\frac{d}{d\theta}\tilde{\gamma}_i$, and $A$ is an arbitrary constant satisfying $0<A<\pi/2$.

Let us then consider the constraint imposed by $\sin\theta=\tilde{b}'_i/\tilde{\gamma'}_i$ for $i=1,2$,
which leads to $\tilde{\gamma'}_i=\tilde{b}'_i/\sin\theta$. To make sure
$\tilde{\gamma}_i$ does not diverge, we can assume that $\tilde{b}_i\left(\theta\right)=f_i\left(-\cos\theta\right)$.
This gives $\tilde{b}'_i\left(\theta\right)=f'_i\left(-\cos\theta\right)\sin\theta$,
so that we have $\tilde{\gamma'}_i(\theta)=f'_i\left(-\cos\theta\right)$. This guarantees that $\dot{\beta}\left(0\right)=\dot{\beta}\left(T\right)=0$ in the time
evolution. The advantage of such a choose is that the Rabi frequencies in Eqs.~(\ref{omegap}) and ~(\ref{omegas}) will no longer diverges
at the initial and final times. Here, to make notations
simpler, we define $m=-\cos\theta$, which is strictly increasing
for $\theta\in\left[0,B\right]$. Thus $\theta=\arccos\left(-m\right)$
is well defined. Therefore, we have $\tilde{b}_i\left(\theta\right)=\tilde{b}_i\left(\theta\left(m\right)\right)\equiv b_i\left(m\right)=f_i\left(m\right)$.
Suppose that $\tilde{\gamma}_i\left(\theta\right)=\tilde{\gamma}_i\left(\theta\left(m\right)\right)\equiv r_i\left(m\right)$, then we have $r'_i\left(m\right)=\frac{d}{dm} r_i \left(m\right) =  \frac{d \theta}{d m}  \frac{d}{d \theta}\tilde{\gamma}_i \left(\theta\right) =  b_i'(m)/\sin \theta = b'_i(m)/ \sqrt{1-m^2}$. The tasks are then transformed into the following forms: find
$b_i\left(m\right)$ and $r_i\left(m\right)$ such that
\begin{eqnarray}    
&&b_{1}\left(-1\right)=0,\;\;b_{1}\left(-\cos B\right)=b_2\left(-\cos B\right)=A,
\nonumber \\
&&b_{2}\left(-1\right)=\pi/2,\;\;\sqrt{1-m^2}=\frac{b_i'(m)}{r_i'(m)}, \label{r}
\end{eqnarray}
where $b'_i\left(m\right)=\frac{d}{dm}b_i\left(m\right)$, $r'_i\left(m\right)=\frac{d}{dm} r_i \left(m\right) = \frac{d}{d \theta} \frac{d \theta}{d m} \tilde{\gamma}_i \left(\theta\right)$ and $ r_1 (-\cos B) = r_2 (-\cos B)$.
Denoting
\begin{eqnarray}
Q_1 \equiv \int_{-1}^{-\cos B}dme^{-i r_1}\left(\frac{i}{\sqrt{1-m^{2}}}-mb_1'\right)
\end{eqnarray}
and
\begin{eqnarray}
Q_2 \equiv\int_{-\cos B}^{-1}dme^{-i r_2}\left(\frac{i}{\sqrt{1-m^{2}}}-mb_2'\right),
\end{eqnarray}
the SES can be rewritten in a compact form as
\begin{eqnarray}    
q_{s}=\left|Q_1+Q_2\right|^{2}.\label{SES}
\end{eqnarray}
From Eq.~~(\ref{r}), we have
\begin{eqnarray}    
r_i \left(m\right)=\int \frac{b'}{\sqrt{1-m^{2}}}dm.
\end{eqnarray}
Let us assume that the ansatz function $b_i\left(m\right)$ takes a linear form, which is the simplest nontrivial function of $m$. That is,
\begin{eqnarray}    
                    b_1(m) = Dm+D,\ \ 0\leq t\leq T/2, \nonumber \\
                   b_2(m) = Fm+F+\pi/{2},\ \ T/2<t\leq T,\label{main1}
\end{eqnarray}
where $D$ is called the phase constant which is to be determined by setting $q_s = 0$, and $F=-2D+\pi+2D\cos B/2\left(-1+\cos B\right)$ is another constant. In this way, the boundary condition of $b_1(-\cos B) = b_2 (-\cos B)$ is satisfied with
$A=-D\cos B+D$. This gives
\begin{eqnarray}    
                  r_1(m) = D\arcsin m,\ \ 0\leq t\leq T/2,
                  \nonumber \\
                  r_2(m) = F\arcsin m+FR-DR,\ \ T/2<t\leq T, \label{main2}
\end{eqnarray}
where $R=\arcsin \left(\cos B\right)$.

To summarize, we have significantly simplified the original boundary conditions by picking some useful functional forms of $\theta$ and $\beta$. We give the final analytic expressions of $b_i$ and $r_i$ in Eqs.~(\ref{main1}) and ~(\ref{main2}), which automatically satisfy the boundary conditions, leaving only the parameters $D$, $B$, and $T$ to be determined. For any given evolution time $T$, we can choose the population parameter $B$, and solve the phase parameter $D$ by setting $q_s$ as a very small value, i.e. $q_s\ll 1$. In general, we cannot prove that the equation $q_s=0$ for the phase parameter $D$ always has a solution, but numerical analysis shows that $q_s$ is at order of magnitude $10^{-2}$, which means that our protocol can suppress the the noise to $1\%$. Then the parameter $\beta$ and $\theta$ for the Hamiltonian $H$ satisfying the boundary equations Eq.~(\ref{boundary}) are well determined when the constant
$D$, $B$, and $T$ are given. Once we have the expressions for $b_i(m)$ and $r_i(m)$, we can calculate the function $\beta(t)$ and $\gamma(t)$ easily, and the Rabi frequencies $\Omega_p$ and $\Omega_s$ can be obtained from Eqs.~(\ref{omegap}) and ~(\ref{omegas}).

The above scheme of finding the parameters for any desired parameters $B$ and $T$ constitutes our main result. Notably, this scheme is not limited to the $100\%$ population transfer from the state $|{1}\rangle$ to $|{3}\rangle$; it can also be applied to arbitrary population transfer, i.e.,  from the state $\cos \beta_0 |{1} \rangle+ \sin \beta_0 |{3}\rangle$ to $\cos \beta_1 |{1}\rangle+ \sin \beta_1 |{3}\rangle$ for arbitrary $\beta_0$ and $\beta_1$. To achieve this arbitrary population transfer, one should set the boundary conditions for $\beta(t)$ as $\beta(0)=\beta_0$ and $\beta(T) =  \beta_1$, which can be easily satisfied by adjusting the parameters $D$ and $F$ in the ansatz function of $b_i$ in Eq.~(\ref{main1}). In the following, we shall see that this solution makes sure that the Rabi frequencies do not diverge in the time evolution, while it can realize a perfect and robust state transfer.

\begin{figure}[t]
\begin{center}
\includegraphics[width=7.8 cm,angle=0]{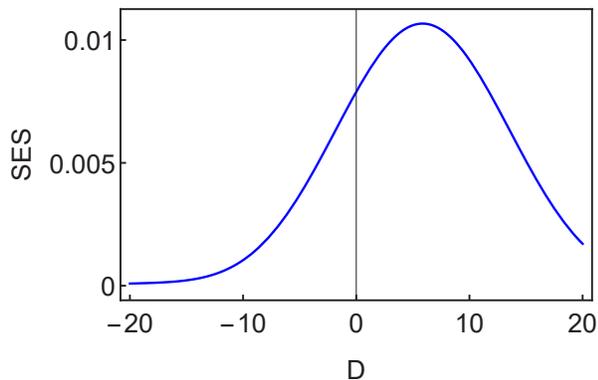}
\caption{SES in Eq. (\ref{SES}) with respect to the parameter $D$. Here we use dimensionless units
with $\hbar=1$, $T=0.1 \mu s$, and $B=\pi/6$. One can see that the condition $q_{s}=0$ is
satisfied when $D\approx-20$. Moreover, the systematic error
sensitivity always takes a very small value compared to $1$ for all process durations with the maximal value being $0.0107$.}\label{fig1}
\end{center}
\end{figure}

\begin{figure}[t]
\begin{center}
\includegraphics[width=9.0 cm,angle=0]{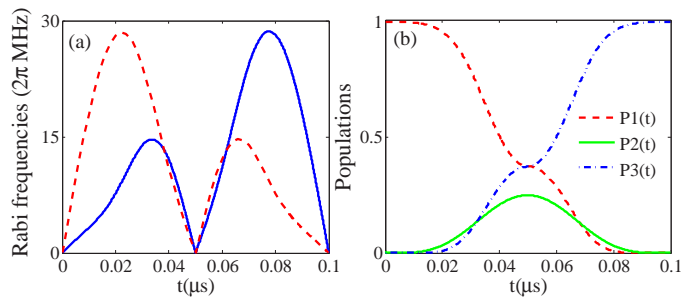}
\caption{ (a) The Rabi frequencies $\Omega_{s}$
(red, dashed line) and $\Omega_{p}$
(blue, solid line) in the optimal STISRAP protocol. (b) Time evolution
of the populations $P_1(t)$ (red, dashed line), $P_2(t)$ (green, solid
line), and $P_3(t)$ (blue, dotted dashed
line) during the population transfer, without considering relaxation. We have used $D=5.85$, $T=0.1\mu s$, and $B=\pi/6$.}\label{fig2}
\end{center}
\end{figure}

We then analyze the value of SES. Substituting the expression for the parameter $\beta$ and $\gamma$ in Eqs.~(\ref{main1}) and ~(\ref{main2}) into the SES expression of Eq.~(\ref{SES}), we get the analytic
form of the systematic error sensitivity, which is plotted in Fig. \ref{fig1}.
The x-coordinate is parameter $D$, and the
y-coordinate represents the systematic error sensitivity, where $B=\pi/6$ and
$T=0.1 \mu s$. Notably, one can find that the SES always takes a very small value ($q_s < 0.015\ll1$)  with the variations of the phase
parameter $D$ and can reach zero when $D\approx-20$. This shows that we have found a family of robust protocols parametrized by the phase parameter $D$ against such a systematic error. Then the
time-dependent Rabi frequencies $\Omega_{p}$ and $\Omega_{s}$ can be obtained, shown in Fig. \ref{fig2} (a),
with maxima $\Omega^m_{p}\approx\Omega^m_{s}=28.50\times2\pi~\rm MHz$.
Here $D=5.85$ is given to get a relatively small values of the maximum of Rabi
frequencies, while the SES is a trivial value below $0.005$. In principle, one can select
a smaller value of $D$ to make the error sensitivity zero, but this may give a large value of Rabi frequencies. The
reasons for choosing such a $D$ are three-manifold: The pulses can possess the same amplitudes and symmetrical shapes
with a a relatively small value of the maximum of Rabi
frequencies, so it is easy to be prepared in experiment; the SES takes a very small value, and thus
the protocol is still robust against the systematic errors; it is also convenient for further comparisons
with other shortcut protocols when the same values of maximum of Rabi
frequencies are designed.

To find the specific form of $H(t)$ in Eq.~(\ref{hamil}),
we solve the Schr\"{o}dinger equation $H(t)|\varphi(t)\rangle =i d|\varphi(t)\rangle/dt$ numerically
by a Runge-Kutta method with an adaptive step, and get the time evolution of the
populations $P_k(t)$ (k=1,2,3) during the population transfer based on the optimal STISRAP protocol, presented in
Fig.~\ref{fig2} (b). The population is defined as $P_k(t)= \langle k|\rho(t)|k\rangle$ with $\rho(t)=|\varphi(t)\rangle\langle\varphi(t)|$
being the final density matrix after time evolution of the population
transfer operation on the initial state $|1\rangle$. One can see that
the time evolutions governed by the Hamiltonian $H(t)$ can achieve a $100\%$ population
transfer from the states $|1\rangle$ to $|3\rangle$, that is, $P_1(0)= 1$ and $P_3(0)= 0$,
and $P_1(T)= 0$ and $P_3(T)= 1$, where $P_i$ denotes the population of the  state $|{i}\rangle$.
The population of the intermediate state $|2\rangle$ is $\sin^{2}\theta$, and it can reach its
maximum $\sin^{2}(\pi/6)=0.25$ when $t=T/2$, as we expected for $B=\pi/6$. Also, a
simple adiabatic approach with Rabi frequencies $\Omega_{{p}(a)}=\Omega^m_{p}\sin\beta$ and $\Omega_{{s}(a)}=\Omega^m_{s}\cos\beta$
needs $T\geq0.56\mu s$ to achieve the same state transfer with $0.9999$ fidelity, so the optimal protocol based on
$H$ is $5.6$ times faster.

\section{The trade-off between the population of the intermediate state and the amplitude of Rabi frequencies}
\label{sec4}

\begin{table*}[t]
\centering \caption{The relation between the maximal population of the intermediate state
($P_{2(peak)}$) and the maximal amplitude of Rabi frequencies ($\Omega_{peak}$)
with the variations of $D$ and $B$.}
\begin{ruledtabular}
\begin{tabular}{ccccccc}
\cline{1-3}
\hline
$D$  & 40.88   &   10.32   &  5.85 &  2.86   & 1.59  \\
   \hline
$B$  &  $\pi/16$   &   $\pi/8$   & $\pi/6$ & $\pi/4$ & $\pi/3$   \\
\hline
$P_{2(peak)}$  & 0.03806     &0.14625   & 0.25  & 0.5  & 0.75   \\
\hline
$\Omega_{peak}$      & 78.246 &   38.5623    & 28.5498  & 18.5065 & 13.6963   \\
\hline
\end{tabular}\label{table1}
\end{ruledtabular}
\end{table*}
In practice, in order to achieve high-fidelity state transfer in three-level quantum systems, one also needs to
minimize the interference from occupation fluctuation of the intermediate level. However, we find that this needs a bigger
amplitude of Rabi frequency, meaning a
lager energy cost. In other words, one cannot simultaneously suppress the population of the intermediate level and
reduce the energy cost. We will show such a trade-off between
these two requirements in quantum state engineering.

Here, we study the relation between the population of the intermediate state and the amplitude of Rabi frequencies
in the state transfer process. These two quantities are two positive factors contributing to the fidelity of the state transfer. To minimize the overall error, one has to balance these two sources of error. Thus, understanding such relation can help us to design an effective way to realize robust state transfer
in different experiments. To show this,
the population of the intermediate level is decreased by choosing different smaller value of $B$
in the optimal shortcut, and then appropriate
Rabi frequencies are designed to achieve state transfer with zero systematic-error sensitivity for a given $D$.

\begin{figure}[t]
\begin{center}
\includegraphics[width=9.0 cm,angle=0]{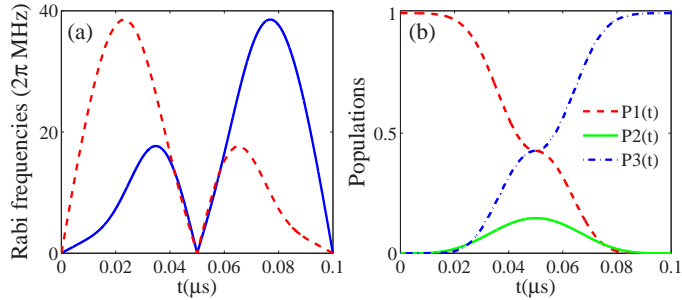}
\caption{ (a) The Rabi frequencies $\Omega_{s}$
(red, dashed line) and $\Omega_{p}$
(blue, solid line) in the optimal STISRAP protocol. (b) Time evolution
of the populations $P_1(t)$ (red, dashed line), $P_2(t)$ (green, solid
line), and $P_3(t)$ (blue, dotted dashed
line) by numerically solving the Schr\"{o}dinger equation in the absence of relaxation. We have used $D=10.32$, $T=0.1\mu s$, and $B=\pi/8$.}\label{fig3}
\end{center}
\end{figure}

First of all, substituting $B=\pi/8$ into Eq.~(\ref{theta}), we set a new function $\theta$. Subsequently,
by plotting the change of the systematic error sensitivity with
the parameter $D$, it is straightforward to find that the SES is always a small value below $0.002$, which means the effect of the noise is suppressed to $2$\textperthousand, suggesting that these family of protocols parameterized by the phase parameter $D$ is generally robust against the systematic error. In fact, an appropriate value of $D$ can be found numerically to make the sensitivity error vanish. Here, $D=10.32$ is chosen to make sure that the maxima of the two Rabi frequencies are the same and they are both relatively small
($\Omega^m_{p}\approx\Omega^m_{s}=38.6\times2\pi~\rm MHz$). Fig. \ref{fig3} (a)
shows the corresponding time evolution of Rabi frequencies
$\Omega_{s}$ (red, dashed line) and $\Omega_{p}$ (blue, solid line). By numerically solving the Schr\"{o}dinger equation
when the initial state is $|1\rangle$, the populations of the three states are depicted in Fig. \ref{fig3} (b).

\begin{figure}[t]
\begin{center}
\includegraphics[width=9.2 cm,angle=0]{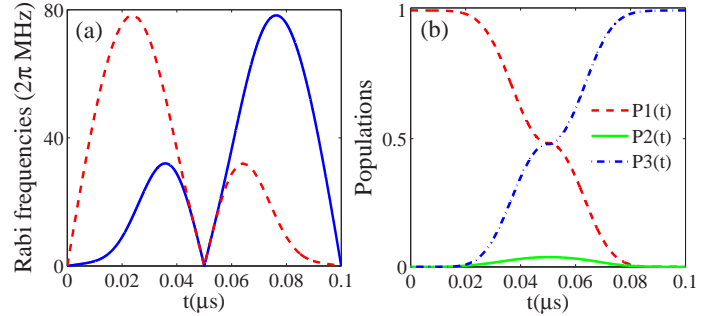}
\caption{(a) The Rabi frequencies $\Omega_{s}$
(red, dashed line) and $\Omega_{p}$
(blue, solid line) in the optimal STISRAP protocol. (b) Time evolution
of the populations $P_1(t)$ (red, dashed line), $P_2(t)$ (green, solid
line), and $P_3(t)$ (blue, dotted dashed
line) by numerically solving the Schr\"{o}dinger equation in the absence of relaxation. We have used $D=40.88$, $T=0.1\mu s$, and $B=\pi/16$.}\label{fig4}
\end{center}
\end{figure}

Secondly, $B=\pi/16$ is chosen to get a smaller population of intermediate state $|2\rangle$. In this case, the
maximum of $P_2$ is $\sin^{2}(\pi/16)\approx0.038$. As a result,
the systematic error sensitivity of the order $10^{-5}$ with respect to the parameter $D$, which means that the systematic error plays a negligible effect
in the three-level population transfer. An equivalent maxima of the Rabi frequencies
requires the value of $D$ to be 40.88. This leads to $\Omega^m_{p}\approx\Omega^m_{s}=78.3\times2\pi~\rm MHz$.
Also, the time evolutions governd by the
Hamiltonian possessing these Rabi frequencies can achieve a perfect population transfer
from initial state $|1\rangle$ to final state $|3\rangle$.
The corresponding time evolution of Rabi frequencies and the populations of the three states are plotted
in Fig. \ref{fig4} (a) and (b), respectively.

From Figs. \ref{fig2}, \ref{fig3}, and \ref{fig4}, the performances of Rabi frequencies
and populations for the three states with variation of time t can be summarized as follows:
(i) The systematic error sensitivity is always small ($q_s\ll 1$) for different $D$. Thus a large range of $D$ can be
prescribed to get a proper $\theta$ and $\beta$, and then the Rabi frequencies are found from Eqs.~(\ref{omegap}) and (\ref{omegas});
(ii) By tuning an appropriate value of $D$, the equivalent maxima of Stokes and pumping pulses can be obtained, and the time
evolution of the quantum state governed by the Hamiltonian with the two pulses can achieve a complete population transfer;
(iii) The smaller the population of the intermediate state is, the larger the amplitudes of Rabi frequencies
are.

One can increase the value of $B$, so that the population of intermediate state $|2\rangle$
is enlarged. Moreover, for $B=\pi$, an appropriate choice of $D$ can lead to a protocol whose SES is exactly zero. Substituting the given $B$ and $D$, the $\theta$ and $\beta$
are determined from the Eqs.~(\ref{main1}) and (\ref{main2}). The corresponding Rabi frequencies also can achieve a $100\%$ state transfer.
The relation between the maximal population of the intermediate state and the maximal amplitude of Rabi frequencies can be summarized in Table.~\ref{table1},
where $P_{2(peak)}$ and $\Omega_{peak}$ represent the maximal population of the intermediate state and the maximal amplitude of Rabi frequencies in
the transfer process, respectively. It shows a close relation between the $P_{2(peak)}$ and $\Omega_{peak}$, namely, one must decrease as the other increases.
This implies that a zero value of $P_{2(peak)}$
is impossible as an infinite energetic cost can not be reached.

\section{Comparison between the optimal invariant-based shortcut and  the original protocol}
\label{sec5}

\begin{figure}[t]
\begin{center}
\includegraphics[width=7.8 cm,angle=0]{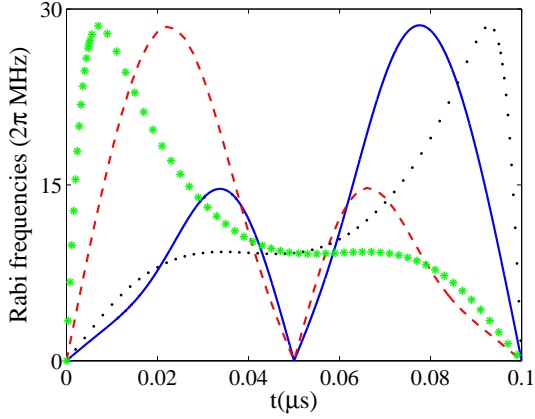}
\caption{ The time evolution of Rabi frequencies $\Omega_{s(o)}$ (green, star line) and $\Omega_{p(o)}$ (black, dotted line)
are obtained from Eqs.~(\ref{omegap}) and ~(\ref{omegas}) by solving the polynomial ans$\ddot{a}$tze $\theta_{o}=\sum_{j=0}^4a_{j}t^{j}$ and $\beta_{o}=\sum_{j=0}^3b_{j}t^{j}$ with boundary conditions in the original invariant-based inverse engineering shortcut, together with the Rabi frequencies $\Omega_{s}$
(red, dashed line) and $\Omega_{p}$ (blue, solid line) in our optimal protocol.}\label{fig5}
\end{center}
\end{figure}
\begin{figure}[t]
\begin{center}
\includegraphics[width=7.8 cm,angle=0]{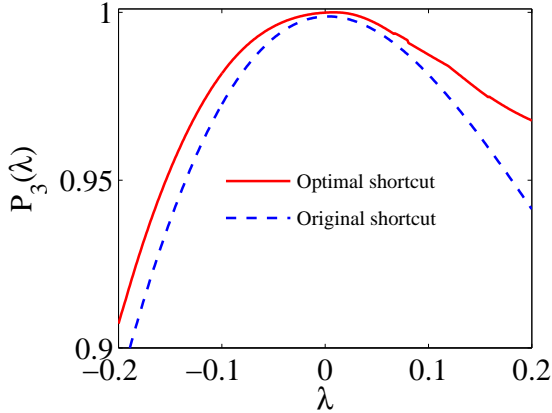}
\caption{ Population $P_3(\lambda)$ at the final time $T=0.1\mu s$ vs systematic
error $\lambda$ by solving numerically the Schr\"{o}dinger equation $\left(H(t)+H'(t)\right)|\psi_{0}(t)\rangle =i d|\psi_{0}(t)\rangle/dt$
when the initial state is $|1\rangle$, based on the optimal invariant-based shortcut in
Fig. \ref{fig2}(a) (zero sensitivity, red, solid line), and the original protocol (nonzero
sensitivity, blue, dashed line).}\label{fig6}
\end{center}
\end{figure}
To show the robustness of our optimal protocol against the systematic error, we compare it with the original
invariant-based inverse engineering in three-level quantum systems proposed by Chen and Muga in
2012~\cite{chen2012}, which shall be referred as the original protocol. Both protocols can achieve the fast and high-fidelity quantum state transfer in a
three-level quantum system along the eigenstate of the invariant $\left|\phi_{0}\left(t\right)\right\rangle$,
in a given time $T$. To avoid the confusion of the symbols, we define the Stokes and pumping pulses in the original
invariant-based shortcut with nonzero sensitivity as
$\Omega_{p(o)}$ and $\Omega_{s(o)}$, respectively, and the corresponding auxiliary angles are $\theta_{o}$ and $\beta_{o}$.
Also, to guarantee a fair comparison,
we let the two protocols have the same maxima of two Rabi frequencies, same population of intermediate state, and same evolution time.
To this end, we should modify some
parameters of the boundary conditions in the original protocol. That is, $\theta_{o}(0)=\theta_{o}(T)=\epsilon$ with $\epsilon=0.035$, which
ensures the maxima of the Rabi frequencies are $\Omega^m_{p(o)}=\Omega^m_{s(o)}\approx28.50\times2\pi~\rm MHz$, as in
the optimal invariant-based shortcut, while other boundary conditions are the same as the Eqs.~(\ref{boun1}) and ~(\ref{boun2}). We also
take $\dot{\beta}_{o}(0)=0$ and $\dot{\beta}_{o}(T)=0$ to make sure that $\Omega_{s(o)}(0)=0$ and $\Omega_{p(o)}(T)=0$.
One set of possible solutions for $\theta_{o}$ and $\beta_{o}$ can be obtained by assuming a polynomial ansatz to interpolate at intermediate
times that $\theta_{o}=\sum_{j=0}^4a_{j}t^{j}$ and $\beta_{o}=\sum_{j=0}^3b_{j}t^{j}$, and then the coefficients can be directly solved in
terms of the above boundary conditions. Finally, the Rabi frequencies $\Omega_{s(o)}$ and $\Omega_{p(o)}$ in the original invariant-based shortcut
can be found from the Eqs.~(\ref{omegap}) and ~(\ref{omegas}). In Fig. \ref{fig5}, we plot
the time evolution of Rabi frequencies $\Omega_{s(o)}$ (green, star line) and $\Omega_{p(o)}$ (black, dotted line)
in the original invariant-based shortcut, together with the optimal ones $\Omega_{s}$
(red, dashed line) and $\Omega_{p}$ (blue, solid line) in our protocol. One can find that the four Rabi frequencies
vary in a different manner, while they take the same maxima in the time evolution.

Now, we can discuss the effect
of systematic errors on the two different protocols. By solving the Schr\"{o}dinger equation
$\left(H(t)+H'(t)\right)|\psi_{0}(t)\rangle =i d|\psi_{0}(t)\rangle/dt$ numerically
with the initial state $|1\rangle$, in both invariant-based shortcuts, the population $P_3(\lambda)$ of state $|3\rangle$ at final time $T$ changes with respect
to parameter of systematic error $\lambda$, demonstrated in Fig. \ref{fig6}. It turns out that the final population in our optimal protocol is always
higher than the original protocol with respect to $\lambda$, showing the robustness of our protocol against systematic error. Also, note that the population of
state $|1\rangle$ in our optimal protocol at the initial time is 1, with no population at the intermediate state at all. This is in contrast to
the original protocol where the population of the intermediate state takes a non-vanishing value, i.e., $\sin^{2}\epsilon$. This shows another advantage of our protocol; the Rabi frequencies are bounded and smooth and thus they can be easily generated in practice.

\section{Summary}
\label{sec6}

In summary, we have proposed a general solution scheme for robust stimulated Raman shortcut-to-adiabatic passage
with invariant-based shortcuts to adiabaticity in three-level quantum systems.
We have inversely engineered a family of optimal Hamiltonians that can suppress the systematic error sensitivity. In particular,
we have studied the relation between the population of the intermediate state and the amplitude of Rabi frequencies,
which shows a trade-off behavior, i.e., the smaller the amplitudes of Rabi frequencies are, the larger population of
the intermediate state is. Moreover, the comparison between the optimal shortcut and the original protocol presents the
robustness of the optimal one against systematic errors in the time evolution. These results suggest that the optimal
stimulated Raman shortcut-to-adiabatic passage indeed provides a general route for robust state transfer in three-level
quantum systems, which is helpful for realizing high-fidelity quantum information processing in different experimental platforms.
\section*{ACKNOWLEDGMENTS}

We would like to thank J. G. Muga for helpful discussions. This work is supported by the National Natural Science Foundation of China under
Grants No. 12004006 and No. 12075001, Anhui Provincial Natural Science Foundation (Grant No. 2008085QA43),
 and Natural Science Foundation of Guangdong Province (2017B030308003) and the Guangdong Innovative
 and Entrepreneurial Research Team Pro- gram (No.2016ZT06D348), and the Science Technology and Innovation
 Commission of Shenzhen Municipality (ZDSYS20170303165926217, JCYJ20170412152620376).

\end{document}